\documentclass[11pt]{article}

\usepackage{graphicx}
\usepackage{amsmath}
\usepackage{amssymb}
\usepackage{algorithmic}
\usepackage{stmaryrd}
\usepackage{hyperref}
\usepackage{moreverb}
\usepackage{nonfloat}
\usepackage{longtable}

\newcommand{\agentlego}[1]{#1_{\textsf{Lego}}}
\newcommand{\agentjason}[1]{#1_{\textsf{Jason}}}

\begin{document}
\title{Implementing Lego Agents Using \emph{Jason}}
\author{Andreas Schmidt Jensen}
\date{1 October 2010}
\maketitle

\begin{abstract}
Since many of the currently available multi-agent frameworks are generally mostly intended for research, it can be difficult to built multi-agent systems using physical robots. In this report I describe a way to combine the multi-agent framework \emph{Jason}, an extended version of the agent-oriented programming language AgentSpeak, with Lego robots to address this problem. By extending parts of the \emph{Jason} reasoning cycle I show how Lego robots are able to complete tasks such as following lines on a floor and communicating to be able to avoid obstacles with minimal amount of coding. The final implementation is a functional extension that is able to built multi-agent systems using Lego agents, however there are some issues that have not been addressed. If the agents are highly dependent on percepts from their sensors, they are required to move quite slowly, because there currently is a high delay in the reasoning cycle, when it is combined with a robot. Overall the system is quite robust and can be used to make simple Lego robots perform tasks of an advanced agent in a multi-agent environment.
\end{abstract}

\clearpage
\tableofcontents
\newpage
\section{Introduction}
Many of the currently available multi-agent frameworks are purely intended for research
purposes and are as such not very suitable for implementing agents in physical robots. 
This paper describes an attempt to built an extension for one such framework that enables
us to use it for implementing physical robots.

The \emph{Jason} framework provides an intuitive and easy way to implement
advanced Multi-Agent Systems. Being highly customizable, it enables
the user to easily represent models of artificial and real scenarios.
However, there is currently no link between real world agents
(such as robots) and software agents (implemented in \emph{Jason}).

The Lego Mindstorms NXT is a robot built in Lego. It includes several
actuators and sensors, which potentially gives unlimited possibilities for
building advanced robots. It comes with a drag-and-drop programming
interface, which makes it impossible to exploit the potential fully, however
by installing the LeJOS Java Virtual Machine on the NXT, the robot is able to
run programs implemented in Java.

A combination of the \emph{Jason} framework with Lego robots having LeJOS installed,
could result in an easy and efficient way to implement real world multi-agent systems.

This paper describes how to implement Lego agents using the \emph{Jason} framework.
While having considered porting the entire framework to the LeJOS JVM, it
seems more reasonable to extend the \emph{Jason} framework to be able to communicate
with Lego robots.

To be able to distinguish between the agent in \emph{Jason} and the physical lego agent, I write
$\agentjason{A}$ to specify that I am talking about the \emph{Jason} part of agent $A$,
and similarly $\agentlego{A}$ when I am refering to the Lego agent of $A$.

\section{Jason}

Instead of changing the general structure of the \emph{Jason} reasoning cycle the idea is to extend parts of it, allowing \emph{Jason} agents to send intentions to the Lego agent and receive percepts.

\begin{figure}[htbp]
\begin{center}
\includegraphics[scale=1]{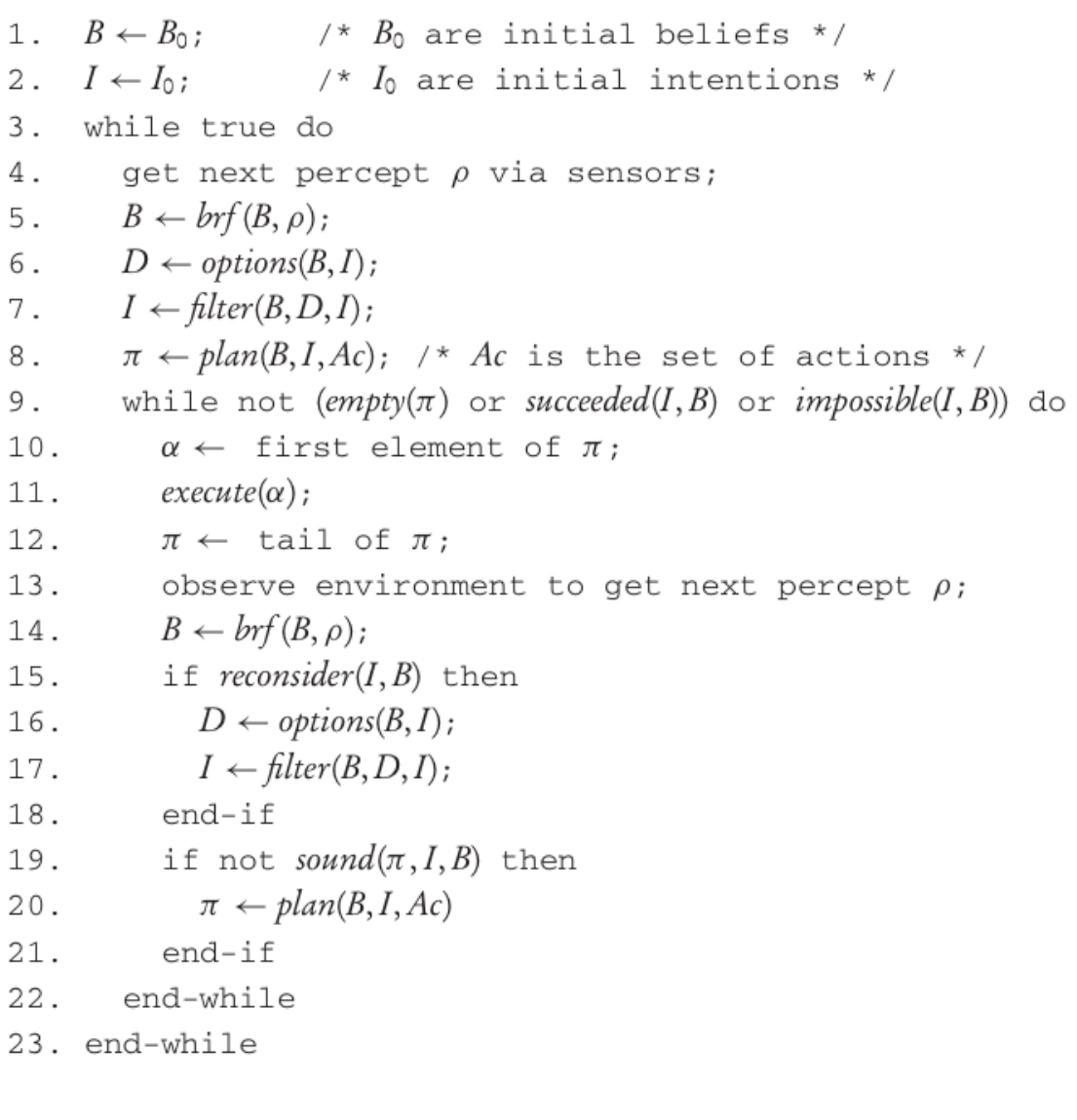}
\caption{The general control loop for a BDI practical reasoning agent\cite{Bordini+2007}.}
\label{fig:control_loop}
\end{center}
\end{figure}

The general control loop for a practical reasoning agent, which is somewhat analogous to the \emph{Jason} reasoning cycle\cite{Bordini+2007}, is seen in figure \ref{fig:control_loop}. While the reasoning cycle is a bit more detailed, the similarities are obvious and it is straightforward to map the steps in the control loop to steps in the reasoning cycle. 

The control loop continuously perceives the environment, updates beliefs, decides what intention to achieve and looks for a plan to achieve it. The plan is then executed until it is empty, has succeeded or is impossible. It should also be noted that the plan is reconsidered often, to ensure that it is still sound. 

To adapt the control loop for real world agents in general, we must look at what parts of the control loop these agents use to interact with the environment. Generally, an agent will be able to perceive the environment and to execute actions. How it perceives and act depends on the sensors and actuators, but generally speaking, this is what a real world agent is able to do.

In order to let Lego agents execute \emph{Jason} plans, the parts of the control loop concerning perceiving the environment and executing plans must be extended. In the general control loop, this means that lines 4, 11 and 13 should be extended. All other lines refers to computations regarding the believes, desires, intentions and plans, all of which are common for agents in general, and it seems reasonable to let these computations happen within \emph{Jason}.

\subsection{Communication}\label{sec:jason_checkingmail}
Being a framework for Multi-Agent Systems, \emph{Jason} naturally allows agents to communicate by sending messages to one another. The Lego agents should also be able to send messages to one another. There are two possibilities for how this can happen:

\begin{itemize}
\item The agents send messages directly to one another. This implies they have knowledge on how to contact each other (the protocol, address, etc). A received message should then be interpreted as a percept and afterwards sent to the \emph{Jason} engine.\\[-8pt]

\item The agents send messages to the \emph{Jason} representations of each another. This only implies knowledge on the name of the agent to contact. The \emph{Jason} engine would then be responsible for everything regarding sending and receiving messages. 
\end{itemize}

The obvious advantage of using the second approach is that no new functionality needs to be implemented. The agents are able to send messages using the available internal actions in \emph{Jason}. In that case, it is already a part of the AgentSpeak language and the agents will readily be able to react to the communication. 

\subsection{Synchronous vs. asynchronous approach}
One thing to keep in mind is the fact that a noticeable amount of delay is going to arise, since the agents will have to wait for the engine to send actions, while the engine will have to wait for the agents to send percepts -- the entire process is synchronous. This means that even though the agents may be able to quickly execute actions and the engine is able to quickly perform a reasoning cycle, the overall performance could be very poor.

It may be more efficient to implement a number of queues, which holds a number of items waiting to be examined and used. The agents holds a queue of actions to perform, while the engine has a queue of percepts to consider. The idea is that if an agent has no items in the queue, it must wait. However, if the engine has not received any percepts, it should keep revising the plans, using the knowledge it currently has available. Eventually, the agent will send new percepts, which would let the engine update the plans to conform with the changes in the environment.

Both approaches have advantages and disadvantages. Using the synchronous approach, the engine and the agents are always up-to-date. No matter what happens in the environment, the engine will know immediately, potentially giving rise to a change in the plan. However, as mentioned, the process is synchronous, meaning that it may prove to perform very poorly. 

In the asynchronous approach, on the other hand, the engine continuously updates the plans and lets the agents execute actions, even though new information is not available yet. This means that the performance could be much better. However, it gives rise to situations where the agent keeps following an impossible plan, because the engine did not yet receive any percepts showing changes in the environment.

\section{Lego Mindstorms NXT}\label{sec:legomindstormsnxt}
The Lego Mindstorms NXT is a toolkit consisting of a set of sensors and motors, a simple computer and a set of Lego  bricks for building robots. Included is a programming language and editor called NXT-G. It allows the user to develop simple applications by a drag-and-drop interface, using loops, branching statements and simple calculations. 

\subsection{Sensors}
The following sensors are included in the Lego Mindstorms NXT kit:

\begin{itemize}
\item The \textbf{touch sensor} is a sensor with a button. The sensor can detect whether or not this button is pressed and can for instance be used to detect whether something has been caught by the arms of a robot. 
\item The \textbf{color sensor} can detect colors. It is tuned to detect standard Lego colors and is able to distinguish between these. 
\item The \textbf{light sensor} is a simpler version of the color sensor, which can distinguish between light and dark areas and could for instance be used to follow a black line on a white surface. 
\item The \textbf{sound sensor} can measure noise level and is also able to recognize sound patterns and identify tone differences.
\item The \textbf{ultrasonic sensor} can measure distances by emitting ultrasonic sound waves. By sensing the time it takes for these sound waves to bounce of an object and return to the sensor, it can measure distances with a precision of 3 centimeters.
\end{itemize}

Other standard and non-standard sensors, such as older sensors can also be used by the Lego Mindstorms NXT, but they will not be discussed here. Refer to appendix \ref{app:extensions} for a description on how to use percepts from these sensors in \emph{Jason}.

\subsubsection{Sampling data from the sensors}
The data provided by the sensors is sent to the extension of the \emph{Jason} framework, so that $\agentjason{A}$ is able to use the values. The samples from a sensor however tend to be somewhat noisy. This means that to successive samples may be very different, even though the environment has not changed.  

The samples are therefore compared to previous values, to see whether they seem reasonable. An easy way to achieve this is to calculate the median of the last $n$ values. This should give a more accurate view of the environment, since extreme values would not be chosen. 

\subsection{Motors}
The kit includes three servo motors, each containing a rotation sensor enabling the user to control the movement of a robot very precisely\footnote{\url{http://mindstorms.lego.com/eng/Overview/Interactive\_Servo\_Motors.aspx}}. It should also be possible to add motors from the previous version of Lego Mindstorms, but this will not be considered here.

A motor can move forward or backwards continuously at a given speed. It can also be controlled to rotate a number of degrees or even to rotate \emph{to} a given degree. The extension to the \emph{Jason} framework will include features allowing agents to move forward, backwards and a rotate number of degrees. This should enable the robots to move around continuously and precisely rotate the motors to adjust to changes in the environment.

The motor also allows for other possibilities and it is also possible to get values from the motor, such as its current speed and angle. This could be used to let the robot increase speed until it has reached some value. I will not go further into details with this, but only use dedicated sensors to perceive. 

Refer to appendix \ref{app:extensions} for a description on how to use the motors in \emph{Jason}.

\section{LeJOS JVM}

The LeJOS JVM is a tiny Java Virtual Machine, which has been ported to the Lego NXT Brick \cite{lejossite}. Because of the limits of the hardware of an NXT brick, the LeJOS JVM is only a subset of standard Java. It does contain many of the standard Java types and classes (lists, input/output support amongst others), but some classes are not included. 

The really useful parts of the LeJOS JVM are the classes giving access to the sensors and motors included with an NXT brick. The motors and each of the standard sensors can be controlled using these classes. The functionality of these classes is what the extension of \emph{Jason} should utilize to control the robots. 

Since \emph{Jason} is not going to be compiled and installed on each brick\footnote{Using standard LeJOS JVM \emph{Jason} cannot even be compiled because of the lack of many classes.}, the two systems should communicate in order to give orders and send percepts. LeJOS includes an API for communicating between a computer and a brick with LeJOS installed, using either bluetooth or USB.

Since applications are excecuted on the brick, the debugging possibilities are quite limited. If an exception is thrown and is not caught, the information about what caused the exception is very sparse. Basically the user is given a  class- and method number. These numbers can then be used to track down where an exception was thrown, by looking at information given by the compiler. LeJOS also gives the possibility of opening a connection to a \emph{remote console}. This forwards all output from the brick to the computer, making it a lot easier to debug applications, since the screen of the brick is quite small. 

\section{Communication}
Communication is a twofold matter, since the system should be able to handle both communication between $\agentjason{A}$ and $\agentlego{A}$ for any agent $A$ as well as communication between two distinct agents, $A$ and $B$. 

\paragraph{$\agentjason{A} \leftrightarrow \agentlego{A}$} 
LeJOS includes a PC API, which enables a computer to communicate with an NXT brick using either bluetooth or USB. While USB is faster than bluetooth, it also highly reduce the usefulness of the robot, since the it will be fastened to the computer by a cable. Using bluetooth, the agent can move much more freely (within the range of the bluetooth connection). Since the communication is mostly going to be small messages containing actions and percepts, a bluetooth connection should not decrease performance noticeably compared to a USB connection.

Some percepts may be more valuable than other but this may also be highly dependent on the problem at hand. Therefore there will not be any sorting in the information perceived, rather will it be send to the computer running \emph{Jason} when available. However, the agents will not add the beliefs immediately to the belief base, since this does not conform with the reasoning cycle. Instead, by overriding the \texttt{perceive()} method, beliefs will only be added at the point in time where the agent is supposed to perceive.

At one step in the reasoning cycle, the agent must perform an action. In \emph{Jason}, one can alter the behaviour of acting by overriding the method \texttt{act()}. Since an actions in this context is something performed by $\agentlego{A}$, the method simply parses the term and sends the corresponding message to $\agentlego{A}$, who will then perform the action immediately. Since LeJOS is a subset of Java, the terms are parsed and converted into a format that is more easily parsed by LeJOS.

Note that even though the use of bluetooth should not decrease communication noticeably, overall slight delays may occur when an action should be performed. First percepts are received and parsed, then the plan is revised and a new intention may be chosen. This may result in choosing a new plan, in which an action must be performed. This action is then parsed and sent to the robot, who performs it. 

\paragraph{\emph{A} $\leftrightarrow$ \emph{B}}
Naturally, the agents in the system must be able to communicate with each other. The question is whether the agents should communicate directly using bluetooth, or using the \emph{Jason} internal actions used for sending messages to one another. 

As discussed in section \ref{sec:jason_checkingmail} there are obvious advantages of using the \emph{Jason} internal actions for communication instead of letting the agents communicate directly. One is that the functionality is already implemented and that by representing each Lego agent as a \emph{Jason} agent, this functionality is readily available. Another advantage is seen when comparing the two alternatives in the case where one agent sends a percept to another agent. Since percepts are used in the \emph{Jason} reasoning cycle to decide upon an intention and find a plan, percepts sent from agent $A$ to agent $B$ will always end up in the percept database of $\agentjason{B}$. 

\begin{table}[htb]
\begin{center}
\begin{tabular}{c|c|c}
\textbf{Step} & \textbf{\emph{Jason} internal actions} & \textbf{Direct communication}\\
\hline
1 & $\agentlego{A} \xrightarrow{Bluetooth} \agentjason{A}$ & $\agentlego{A} \xrightarrow{Bluetooth} \agentlego{B}$ \\
2 & $\agentjason{A} \xrightarrow{Internal} \agentjason{B}$ & $\agentlego{B} \xrightarrow{Bluetooth} \agentjason{B}$ \\
3 & & ($\agentlego{A} \xrightarrow{Bluetooth} \agentjason{A})$ 
\end{tabular}
\end{center}
\caption{Comparison of communication methods when sharing percepts.}
\label{tab:communication_comparison}
\end{table}

Table \ref{tab:communication_comparison} shows the communication that will happen between the agents (of both kinds), when a percept is sent. The third message sent during the direct communication is not exactly relevant for the process it self, however, whenever an agent $\agentlego{A}$ perceives something, this is automatically sent to the  corresponding agent $\agentjason{A}$. 

The differences should be clear. By using direct communication, one sends three messages using bluetooth, while the use of internal actions enables one to limit the number of bluetooth messages to one. Sending a message internally in \emph{Jason} is assumed to be faster than sending a message using bluetooth. It is therefore straightforward to conclude that using internal actions is much more efficient than using bluetooth communication. 

\section{Case studies}
I have developed a few applications entirely in \emph{Jason} to show the abilities of the Lego-\emph{Jason} framework. First I discuss a classical robot example, and then I try to let two agents cooperate in order to perform a simple task.

\subsection{Line Follower}
The line following robot is basically a robot, which given an environment with a marked path (i.e. a line), is able to follow exactly that path. It requires the use of one or more sensors, which is able to perceive what is beneath the robot. I have implemented a version using two Lego light sensors. By putting these sensors close to the floor, the robot is able to see whether it is moving on a line which is distinguishable from the rest of the floor (e.g. black line on white floor).  

The two sensors are placed so that if the agent is on the line, both sensors will show bright values. Once the path is turning, one of the sensors will begin to read darker values, something the agent should react to.

The implementation basically consists of the following three lines (plus a few rules, which could have been explicitly written in each plan\footnote{The entire implementation is included in appendix \ref{app:source}}):

\footnotesize
\begin{verbatim}
+!move : light(S1, V1) & light(S2, V2) 
         & distinct_sensors(S1, S2) & on_line(V1, V2)
         <- forward([a,b],[60,60]); !!move.
+!move : light(1, V1) & light(2, V2) & turning(V1,V2)
         <- rotate([a,b],[-30,15]); !!move.
+!move : light(1, V1) & light(2, V2) & turning(V2,V1)
         <- rotate([a,b],[15, -30]); !!move.
\end{verbatim}
\normalsize

The agent has a goal \texttt{!move}, which it at all times will pursue. The first line says that if it has perceived light values from two distinct sensors, and they indicate that the robot is on the line, then it should move forward. The second and third line are the cases where the agent is about to leave the line and must turn to avoid this. The two cases corresponds to the line turning left and right, respectively. 

It must be noted that in order to successfully follow lines, the robot must move very slowly forward, so that it is able to react to percepts before it is too late (i.e. before it has moved completely away from the line). This imples that communcation is a greater bottleneck than initially suspected.

The implementation assumes that the agent initially is placed on the line, and the result may otherwise be unpredictable. However, given that assumption, the entire implementation can be done in three lines.

\subsection{Communication}
In the second case study I consider two agents: One with a light sensor, an ultrasonic sensor and two motors -- the ``obstacle finder'' and one with only a light sensor and two motors -- the ``blind agent''. The agents are placed in a very simple environment resembling a pedestrian crossing, i.e. alternating white and black bars, on which is placed an obstacle that the agents must avoid when crossing the environment. Both agents know how to avoid the obstacle, when they find it, but none of them initially know where it is. 

The problem is to ensure that both agents successfully cross the path without touching the obstacle. The obstacle finder is able to detect the obstacle by using its ultrasonic sensor and can therefore fairly easily avoid it. The blind agent can only detect how many black bars it has passed. The idea is therefore to let the obstacle finder send a message containing the number of black bars to pass, before the agent must avoid the obstacle. This information can then be used by the blind agent to successfully complete the task.

This problem is easily solved using \emph{Jason} when noted that the agents have a goal to avoid the obstacle and that they should react to changes in the environment. The implementation is found in appendix \ref{app:source}. Below is shown code snippets that show how the agents cooperate:

\subsubsection*{Obstacle Finder}
\footnotesize
\begin{verbatimtab}[4]
+light(_, X)[source(percept)] : goal(search) & on_bar(X) & last_color(white) 
		<-	-+last_color(black); ?bars_passed(N); 
			BarsPassed = N + 1; -+bars_passed(BarsPassed).
+light(_, X)[source(percept)]
		:	goal(search) & not on_bar(X) & last_color(black)
		<-	-+last_color(white).
+obstacle(_, X)[source(percept)]
		: 	goal(search) & X < 15 
		<- 	-+goal(avoid); ?bars_passed(N);				 
			.send(blindagent, tell, obstacle_after(N)); !!avoid.
\end{verbatimtab}
\normalsize

Basically, the agent has a counter incrementing each time it moves over a black bar. When an obstacle is observed near the agent, it tells the blind agent at which bar the obstacle was found. Then it begins to pursue the goal \texttt{!avoid}.

\subsubsection*{Blind Agent}
\footnotesize
\begin{verbatimtab}[4]
+light(_, X)[source(percept)] 
				:	goal(search) & on_bar(X) & last_color(white) 
				<-	-+last_color(black); ?bars_passed(N); 
					BarsPassed = N + 1; -+bars_passed(BarsPassed).
+light(_, X)[source(percept)]
				:	goal(search) & not on_bar(X) & last_color(black)
				<-	-+last_color(white).
+bars_passed(N)	:	obstacle_after(N) & goal(search)
				<-	-+goal(avoid); !!avoid.
\end{verbatimtab}
\normalsize

The agent uses the same approach to count the number of bars passed, since it also has a light sensor. It pursues the goal \texttt{!avoid}, whenever it perceives that it is near the obstacle (i.e. has passed the same number of bars as the obstacle finder).

The program could also be extended to include the case where the blind agent does not know how to avoid the obstacle. This can be achieved by letting the obstacle finder send the plan for avoiding, using the internal action \texttt{tellHow}.

This case study shows that the agents are able to cooperate and that messaging is actually working quite efficiently. It must be noted that information in the message should be reacted to immediately after reception, so in other examples, where fast reaction to messages is necessary, results may not be as promising.

\section{Future thoughts}
As was shown in the case studies, the system is able to map $\agentjason{A}$ to $\agentlego{A}$. This means that given an implementation of a system in \emph{Jason}, a group of Lego robots would be able to act as the agents implemented in \emph{Jason}. However, some parts of the system are not as efficient as they could be, and there are still plenty of possibilities for extending the system further. In the following I will discuss some of these issues and possibilities.

\paragraph{Performance}
It was shown that performance is not optimal, since in order to give $\agentjason{A}$ enough time to react on a percept sent by $\agentlego{A}$, the latter is forced to act very slowly. In order to make the system really useful, it would be necessary to improve performance dramatically. 

\paragraph{Navigation}
The multi-agent case study revealed one obvious problem with the current solution: the agent has no idea where it is, and has therefore little possibility of navigating. Actually being able to have knowledge of the environment could make it possible for the agent to know where it is and enable it to navigate from $(x_1, y_1)$ to $(x_2, y_2)$ in a world using a navigation algorithm.

Gaining knowledge of the environment is possible by using the ultrasonic sensor, since it is able to detect obstacles, such as walls. The problem is that this knowledge cannot directly be used to create a map, since the robot does not know where it is with regards to that obstacle. It knows that if it has detected an obstacle $x$ centimeters away, then it is $x$ centimeters from an obstacle. However, it has no starting point to combine this knowledge with, since it does not exactly know how far it has moved, before the obstacle was detected.

A technique called \emph{odometry} uses data from the movement of the actuators (motors) to estimate how far the robot has moved over time. Since LeJOS is able to give some information regarding the motors, it may be possible to use this to generate a world model. In \cite{Shatkay+1997} odometry is used to create such map, and it could be very useful to implement such feature in \emph{Jason} and Lego Mindstorms. 

By sharing such map between all agents in the environment, it would be possible for the agents to find and assist each other pursuing their goals.

\paragraph{Sensors}
The system is currently able to handle the four standard types of sensors: Touch, light, sound and ultrasonic sensor. The LeJOS JVM includes support for many other types of sensors, including the sensors from the first version of Lego Mindstorms. As a future addition, it could be relevant to add support for these sensors as well. 

\paragraph{Actions} 
An action is something performed by $\agentlego{A}$. This means that whenever a term in the implementation is not a \emph{Jason} primitive or an internal action, it will be regarded as an action, and therefore be parsed and sent to the Lego agent. 

Section \ref{sec:legomindstormsnxt} describes the actions that the system is able to perform. The LeJOS JVM gives a few more possibilities, that could be supported by \emph{Jason} as well, by implementing the actions in the parser. However, the big issue is that the robot only has one type of actuator, the standard NXT motor. By adding support for other actuators, such as motors with less precision but with more power, new possibilities could become available.

\section{Conclusion}
The main goal of this project has been to implement an extension for the \emph{Jason} multi-agent system that allows one to write applications executed by Lego robots. This would enable the programmer to write multi-agent systems in a logic programming language and even execute these applications on a real robot. 

I have examined the \emph{Jason} reasoning cycle to realize how to communicate with a robot in terms of sending actions and receiving percepts. This has resulted in an implementation, which does not alter \emph{Jason}, but only adds functionality to allow communication with Lego robots. 

The extension to \emph{Jason} is functional, and enables Lego robots to perform actions sent from \emph{Jason}. Furthermore, it was shown that the agent is able to react to the environment by using sensors. It must be noted that while the case studies showed that the agents can perform tasks, they cannot do this fast, since delay must be taken into account when sending percepts, choosing an intention and sending an action back. 

Performance-wise, much could be improved, however the robots are easily programmed using \emph{Jason}, since the language naturally expresses how agents react to the environment and specify how they act to these percepts.

\newpage

\appendix

\section{Extensions of the \emph{Jason} syntax}\label{app:extensions}
In this appendix I briefly describe the actions and percepts that can be used in the LEGO extension of \emph{Jason}. The design of a \emph{Jason}-project file \texttt{project.mas2j} for LEGO projects is also described.

\subsection{Actions}
\begin{longtable}{lp{0.6\linewidth}}
\textbf{forward(Motors, Speed)} & Moves a list of motors forward at given speed. \\[10pt]
& \textbf{Motors:} List of motors. \\
& \textbf{Speed:} List of speed, corresponding to the motor of the same index in the other list. \\

\multicolumn{2}{p{\linewidth}}{
Examples: 
\begin{itemize}
\item \texttt{forward([a,b],[300,300])}: The motors A and B will move forward at speed 300.
\end{itemize}
}\\[20pt]

\hline\\[-8pt]

\textbf{backward(Motors, Speed)} & Moves a list of motors backward at given speed. \\[10pt]
& \textbf{Motors:} List of motors. \\
& \textbf{Speed:} List of speed, corresponding to the motor of the same index in the other list. \\

\multicolumn{2}{p{\linewidth}}{
Examples: 
\begin{itemize}
\item \texttt{backward([a,b],[300,300])}: The motors A and B will move backward at speed 300.
\end{itemize}
}\\[20pt]

\hline\\[-8pt]

\textbf{rotate(Motors, Speed)} & Rotates a list of motors backward a given angle. \\[10pt]
& \textbf{Motors:} List of motors. \\
& \textbf{Speed:} List of angles, corresponding to the motor of the same index in the other list. \\

\multicolumn{2}{p{\linewidth}}{
Examples: 
\begin{itemize}
\item \texttt{rotate([a,b],[300,300])}: The motors A and B will rotate 300 degrees.
\item \texttt{rotate([a,b],[100,-100])}: Motor A will rotate 100 degrees, and motor B will rotate -100 degrees.
\end{itemize}
}\\[20pt]

\hline\\[-8pt]

\textbf{reverse(Motors)} & Reverse the direction of a list of motors. \\[10pt]
& \textbf{Motors:} List of motors. \\

\multicolumn{2}{p{\linewidth}}{
Examples: 
\begin{itemize}
\item \texttt{reverse([a,b])}: The motors A and B will reverse their direction. This happens individually. If A was moving forward and B was moving backwards, A will now move backwards and B forwards.
\end{itemize}
}\\[20pt]

\hline\\[-8pt]

\textbf{speed(Motors,Speed)} & Sets the speed for each motor in the list. \\[10pt]
& \textbf{Motors:} List of motors. \\
& \textbf{Speed:} List of speed, corresponding to the motor of the same index in the other list. \\

\multicolumn{2}{p{\linewidth}}{
Examples: 
\begin{itemize}
\item \texttt{speed([a,b],[200,300])}: Sets the speed of A to 200 and the speed of B to 300.
\end{itemize}
}\\[20pt]

\hline\\[-8pt]

\textbf{stop(Motors)} & Stops the motors in the list. \\[10pt]
& \textbf{Motors:} List of motors. \\

\multicolumn{2}{p{\linewidth}}{
Examples: 
\begin{itemize}
\item \texttt{stop([a,b])}: Stops motor A and B.
\end{itemize}
}\\[20pt]

\hline\\[-8pt]

\textbf{block(Blocking)} & Sets whether the rotation command should return control to the thread immediately. \\[10pt]
& \textbf{Blocking:} Boolean value specifying whether rotation is blocking or control should be return immediately. \\

\multicolumn{2}{p{\linewidth}}{
Examples: 
\begin{itemize}
\item \texttt{block(true)}: Blocks further actions of the robot until a rotation is done. This has no immediate effect, but affects how \texttt{rotate(Motors, Speed)} works. 
\end{itemize}
}\\[20pt]

\hline\\[-8pt]

\textbf{exit} & Tells the (Lego) agent to shut down. \\[10pt]

\multicolumn{2}{p{\linewidth}}{
Examples: 
\begin{itemize}
\item \texttt{exit}
\end{itemize}
}\\[20pt]

\end{longtable}

\subsection{Percepts}
\begin{longtable}{lp{0.6\linewidth}}
\textbf{light(Port, Value)} & Percept from a light sensor. \\[10pt]
& \textbf{Port:} Port on which the sensor is connected. Can be used to distinguish between several sensors of the same kind. \\
& \textbf{Value:} Light value read from the sensor. \\

\multicolumn{2}{p{\linewidth}}{
Examples: 
\begin{itemize}
\item \texttt{light(1,360)}: The light sensor on port 1 has read a value of 360.
\end{itemize}
}\\[20pt]

\hline\\[-8pt]

\textbf{obstacle(Port, Value)} & Percept from an ultrasonic sensor. \\[10pt]
& \textbf{Port:} Port on which the sensor is connected. Can be used to distinguish between several sensors of the same kind. \\
& \textbf{Value:} Distance to obstacle read from the sensor (max = 255). \\

\multicolumn{2}{p{\linewidth}}{
Examples: 
\begin{itemize}
\item \texttt{obstacle(1,40)}: The ultrasonic sensor on port 1 has found an obstacle 40 centimeters away.
\end{itemize}
}\\[20pt]

\hline\\[-8pt]

\textbf{touching(Port, Value)} & Percept from a touch sensor. \\[10pt]
& \textbf{Port:} Port on which the sensor is connected. Can be used to distinguish between several sensors of the same kind. \\
& \textbf{Value:} The value is either true or false, corresponding to whether or not something is touching the button of the sensor. \\

\multicolumn{2}{p{\linewidth}}{
Examples: 
\begin{itemize}
\item \texttt{touching(1, true)}: The touch sensor on port 1 has detected a touch.
\item \texttt{touching(1, false)}: The touch sensor on port 1 has not detected a touch.
\end{itemize}
}\\[20pt]

\hline\\[-8pt]

\textbf{sound(Port, Value)} & Percept from a sound sensor. \\[10pt]
& \textbf{Port:} Port on which the sensor is connected. Can be used to distinguish between several sensors of the same kind. \\
& \textbf{Value:} Sound value read from the sensor. \\

\multicolumn{2}{p{\linewidth}}{
Examples: 
\begin{itemize}
\item \texttt{sound(1,55)}: The sound sensor on port 1 has read a value of 55.
\end{itemize}
}\\[20pt]

\end{longtable}

\subsection{Project file}
I will not go into details with how a general project file looks. I will only describe which classes an agent must use for their architecture and belief base, and which parameters are available and must be specified, to allow \emph{Jason} to connect the the Lego robots.

An agent in the file \texttt{project.mas2j} must look like this:

\footnotesize
\begin{verbatimtab}[4]
agentname agentsource.asl 
	[btname="BTNAME", btaddress="BTADDRESS", 
	motora="t/f", motorb="t/f", motorc="t/f", 
	sensor1="...", sensor2="...", sensor3="...", 
	sensor4="...", sleep="..."]
	agentArchClass arch.LEGOAgArchitecture
	beliefBaseClass agent.UniqueBelsBB("light(port,_)","sound(port,_)",
								"obstacle(port,_)","touching(port,_)");
\end{verbatimtab}
\normalsize

The \texttt{agentArchClass} and \texttt{beliefBaseClass} must be set to the specified classes. The agent architecture takes care of all communication between $\agentjason{A}$ and $\agentlego{A}$. The belief base makes sure that at any time, we can have at most one percept from a single sensor. This is done to make sure that the belief base does not increase exponentially in size. Below is shown the parameters that can be passed to the agent. Parameters in bold are mandatory. 

\begin{longtable}{l||p{0.7\linewidth}}
\textbf{Key} & \textbf{Value} \\
\hline
\hline
\textbf{btname} & Name of the NXT brick \\
\hline
\textbf{btaddress} & Bluetooth address in the format 12:34:56:78:90:AB \\
\hline
\textbf{motor\{a,b,c\}} & true or false, depending on whether the motor is connected to the brick. \\
\hline
\textbf{sensor\{1,2,3,4\}} & the name of the sensor connected or ``none''. The name can be one of \{touch,light,sound,ultrasonic\}.\\
\hline
sleep & Milliseconds a sensor should sleep between sampling. Default: 50 ms.\\
\end{longtable}

\section{Case studies: Source code}\label{app:source}
\subsection{Line Follower}
\subsubsection*{linefollower.asl}
\footnotesize
\begin{verbatim}
// Linefollower agent for the Lego Mindstorms NXT
// Uses two sensors to detect whether it is on the line.
// The agent assumes that it is initially on a line 
// (i.e. both sensors are showing bright light).
// It may therefore not work if not placed on a line 

/* Initial beliefs and rules */
// the sensors must be different, otherwise the light value
// from one sensor will be compared to it self
distinct_sensors(Sensor1, Sensor2) :- Sensor1 \== Sensor2.  
// the robot is on the line if both sensors are showing a 
// bright value (currently a value greater than 350)
on_line(Value1, Value2) :- Value1 >= 350 & Value2 >= 350.
// the robot should turn if one value is dark and the other is bright 
// (i.e. the line is turning)
turning(Value1, Value2) :- Value1 < 350 & Value2 >= 350.

/* Initial goals */
!move.

/* Plans */
// move forward if both sensors are sampling bright values
+!move : light(S1, V1) & light(S2, V2) 
         & distinct_sensors(S1, S2) & on_line(V1, V2)
         <- forward([a,b],[60,60]); !!move.

// turning
+!move : light(1, V1) & light(2, V2) & turning(V1,V2)
         <- rotate([a,b],[-30,15]); !!move.
+!move : light(1, V1) & light(2, V2) & turning(V2,V1)
         <- rotate([a,b],[15, -30]); !!move.
		
// if nothing has been perceived yet, wait, then reintroduce goal.
+!move : not light(_,_)
         <- .wait("+light(_,_)"); !move.
			
// upon failure, start from scratch and reintroduce.
-!move <-	.drop_all_desires; .abolish(light(_,_)); !!move.
\end{verbatim}
\normalsize

\subsection{Communication}
\subsubsection*{obstaclefinder.asl}
\footnotesize
\begin{verbatimtab}[4]
// Agent for finding obstacles and avoiding them.
// When an obstacle is found, the agent sends information
// to a blind agent about how to avoid the obstacle.

/* Initial beliefs and rules */
// How many bars have the agent currently passed
bars_passed(0).

// Assume we alternate between black and white bars
last_color(white).

// intially the agent searches for an obstacle
goal(search).

// The agent is on a bar if the sensors is reading a 
// value less than 350
on_bar(Value) :- Value < 350.

/* Initial goals */
!init.

/* Plans */
+!init	<-	block(true); !!move.

// Increment bar-counter when on a black bar and last bar was white. 
// Also make mental note that the agent is on a black bar now.
+light(_, X)[source(percept)]
		:	goal(search) & on_bar(X) & last_color(white) 
		<-	-+last_color(black); ?bars_passed(N); 
			BarsPassed = N + 1; -+bars_passed(BarsPassed).
// Make mental note that now the agent is on a white bar.
+light(_, X)[source(percept)]
		:	goal(search) & not on_bar(X) & last_color(black)
		<-	-+last_color(white).
// when an obstacle is near -- tell the blind agent and avoid!
+obstacle(_, X)[source(percept)]
		: 	goal(search) & X < 15 
		<- 	-+goal(avoid); ?bars_passed(N);				 
			.send(blindagent, tell, obstacle_after(N));
			!!avoid.

// Avoiding the obstacle. It is assumed that an obstacle 
// can only be of one size.
 +!avoid <-	stop([a,b]); speed([a,b],[300,300]);
 			rotate([a,b],[-200,200]); rotate([a,b],[400,400]); 
 			rotate([a,b],[200,-200]); rotate([a,b],[800,800]); 
 			rotate([a,b],[200,-200]); rotate([a,b],[400,400]); 
 			rotate([a,b],[-200,200]); !!move.
 				
 // approach the obstacle slowly.
 +!move	<-	forward([a,b],[60,60]).
\end{verbatimtab}
\normalsize

\subsubsection*{blindagent.asl}
\footnotesize
\begin{verbatimtab}[4]
// Blind agent that can avoid obstacles, 
// when it is told where they are.
/* Initial beliefs and rules */
// How many bars have the agent currently passed
bars_passed(0).

// Assume we alternate between black and white bars
last_color(white).

// intially the agent searches for an obstacle
goal(search).

// The agent is on a bar if the sensors is reading a value less than 350
on_bar(Value) :- Value < 400.

/* Initial goals */
!init.

/* Plans */
+!init	<-	block(true); !!move.

// Increment bar-counter when on a black bar and last bar was white. 
// Also make mental note that the agent is on a black bar now.
+light(_, X)[source(percept)]
		:	goal(search) & on_bar(X) & last_color(white) 
		<-	-+last_color(black); ?bars_passed(N); 
			BarsPassed = N + 1; -+bars_passed(BarsPassed).
// Make mental note that now the agent is on a white bar.
+light(_, X)[source(percept)]
		:	goal(search) & not on_bar(X) & last_color(black)
		<-	-+last_color(white).
// when the amount of bars passed means the obstacle is near -- avoid!
+bars_passed(N)
		:	obstacle_after(N) & goal(search)
		<-	-+goal(avoid); !!avoid.
// Avoiding the obstacle. It is assumed that an obstacle can only be of one size.
 +!avoid <-	stop([a,b]); speed([a,b],[300,300]);
			rotate([a,b],[-200,200]); rotate([a,b],[400,400]); 
			rotate([a,b],[200,-200]); rotate([a,b],[800,800]); 
			rotate([a,b],[200,-200]); rotate([a,b],[400,400]); 
			rotate([a,b],[-200,200]); !!move.
 				
 // approach the obstacle slowly.
 +!move	<-	forward([a,b],[60,60]).
\end{verbatimtab}

\section*{Acknowledgement}
I thank associate professor J{\o}rgen Villadsen, DTU Informatics, Denmark.

\section*{Source code}
The source code for the system is available here:
\begin{center}
\url{http://imm.dtu.dk/~jv/LEGO-Jason-NXT}
\end{center}

\end{document}